\documentclass[twocolumn,preprintnumbers,amsmath,amssymb]{revtex4-1}
\usepackage{epsfig}
\usepackage{color}
\usepackage{amsmath}
\usepackage{multirow}

\begin{document}

\title{The kinetics of the ice-water interface from ab initio machine learning simulations}

\author{P. Montero de Hijes$^{1}$, S. Romano$^1$, A. Gorfer$^{1,2}$, and C. Dellago$^{1,*}$ }
\affiliation{$^1$Faculty of Physics, University of Vienna, A-1090
Vienna, Austria}
\affiliation{$^2$Department of Lithospheric Research, University of Vienna, Josef-Holaubuek-Platz 2, 1090, Vienna, Austria}

\begin{abstract}

  Molecular simulations employing empiric force fields have provided valuable knowledge about the ice growth process in the last
 decade. The development
 of novel computational techniques allows us
 to study this process, which requires long simulations
  of relatively large systems, with ab initio accuracy. In this work, we use a neural-network potential for water trained on the Revised Perdew–Burke–Ernzerhof functional to describe the kinetics of the
  ice-water interface. We study both ice melting and growth
   processes. Our results for the ice growth
   rate are in reasonable agreement with previous experiments
    and simulations. We find that the 
     kinetics of ice melting presents a different behavior (monotonic)
     than that of ice growth (non-monotonic). In particular, a maximum in the ice
      growth rate of 6.5 $\text{\AA}$/ns is found at 14 K of supercooling.
       The effect of the
       surface structure is explored by investigating 
        the basal and primary and secondary prismatic facets. 
     We use the Wilson-Frenkel relation 
     to explain these results in terms of the mobility of molecules
      and the thermodynamic driving force.
        Moreover, we study the effect of pressure by complementing the standard isobar with simulations at negative pressure (-1000 bar) and at high pressure (2000 bar). We find that prismatic
         facets grow faster than the basal one, and
          that pressure does not play an important 
          role when the speed of the interface
           is considered as a function of the difference between the melting  temperature and the actual one, i.e. to the degree
             of either supercooling or overheating. 
            % We also 
            % approximately estimate the
            %  interfacial free energy at coexistence from the Turnbull relation
            %  finding a good agreement with previous works. 

\end{abstract}

\maketitle
$^*$christoph.dellago@univie.ac.at

\section{Introduction}

 Among all possible phase transitions, 
 the one taking place between
  ice and water is particularly relevant
  in a broad range of fields, 
                  including climate modelling\cite{mcguffie2014climate}, cryobiology\cite{mazur1970cryobiology}, 
                  and aerospace technology\cite{huang2019survey}.  This process
   starts when the original phase, let's assume water, 
   is metastable with
    respect to ice. Then, the transition needs to
     wait until an  ice critical nucleus 
      is formed within the metastable water.
         In pure      substances, this activated
          process  is called homogeneous
         nucleation whereas in the presence of 
      surfaces 
       or impurities, it
        is called
        heterogeneous nucleation\cite{kashchiev2000nucleation,kashchiev2003nucleation}. 
     Homogeneous nucleation in supercooled water
       has been  widely studied\cite{sanz2013homogeneous,espinosa2014homogeneous,espinosa2016time,espinosa2016interfacial,bianco2021anomalous,piaggi2022}. However, the homogeneous nucleation
        of water in overheated ice
        has been scarcely explored.
        The reason is
         that ice, in nature, usually 
          presents a metastable liquid layer in its surface even below the melting point \cite{vega2006absence,conde2008thickness,nagata2019surface,sanchez2017experimental,slater2019surface}.
          Thus, when temperature goes just above the melting,
           the critical size has  already been largely bypassed
           by the liquid layer 
           and the melting occurs immediately from the surface.
           Nevertheless, under special
            conditions, overheating is possible and
             ice melts from the bulk\cite{schmeisser2007maximum, moritz2021microscopic,roedder1967metastable}.
             Therefore, exploring the overheated
              ice melting is important too. 
             Once the critical size is overcome,
            the stable phase keeps growing. In this article,
             we are concerned with the mechanism and kinetics of the
              growth process. \\
            
    The kinetics of the ice-water interface, 
     has already been studied  in the supercooled
      regime\cite{nada1996anisotropic,rozmanov2011temperature,rozmanov2012anisotropy,espinosa2016time,montero2019ice,addula2022kinetic}, 
      by means of empirical force fields
       including the TIP4P\cite{jorgensen1983comparison}, TIP4P/2005\cite{abascal2005general}, TIP4P/Ice\cite{abascal2005potential},
        and mW\cite{molinero2009water}. 
      The molecular interactions in these models
       are represented by empirical relations 
     with parameters which are optimized in order to 
       reproduce as accurately as possible different
        properties of water.  This allows for high
        computational efficiency.
            Despite the simplicity
             of the models, the accuracy of TIP4P/2005
              and TIP4P/Ice is remarkable\cite{abascal2005general,abascal2005potential}.
          However, they cannot capture the effect
           of vibrational motion, or describe
            chemical reactions by construction. When an interface
             is present, these factors may be
              quite relevant.  Ab initio molecular
             dynamics simulations (AIMD)
             \cite{car1985unified}, on the other hand, can describe
              reactions and incorporate any
               kind of motion in the molecules.
               Indeed, the interactions between
                  molecules 
                 arise 
                  from first principle's calculations 
                  and this provides
                 a complete quantum-mechanical description
                 for the system. 
                 However, the cost
                  of these simulations make them
                   prohibitive to study the kinetics
                    of the ice-water interface
                    which requires relatively large systems
                    and long simulations.\\
            
    The development of novel computational techniques
    in the field of machine learning has opened the door
     to bridge the computational efficiency of
      empiric force fields with the quantum-mechanical accuracy
       of ab initio simulations. One method that 
       has gained popularity is the neural-network potential (NNP). 
        In this approach,
         a neural-network is trained on energies
          and forces obtained from AIMD calculations on
           short timescales and relatively small  systems. Then,
           the neural-network is used in classical
           molecular dynamics (MD) over longer 
           timescales and larger systems. 
           This approach has been used for equilibrium
             simulations of
             water in the vapor, liquid, and solid state\cite{morawietz2013density,morawietz2016van, wohlfahrt2020ab, cheng2019ab,zhang2021phase}.
            Recently, Piaggi et al.\cite{piaggi2022} 
            investigated 
             homogeneous ice nucleation 
             by means of
               an NNP trained
                on the SCAN functional.
                In our work, we employ NNPs to study ice growth
                 and melting.\\

    A theoretical approach to explain  the growth of ice is 
    the Wilson-Frenkel framework\cite{montero2019ice,garcia2006melting} which has been applied to a diverse
                   variety of crystals\cite{freitas2020uncovering,yang2021crystal,ripoll1996theoretical}. The main
                    idea is that the growth
                    is determined by the mobility of molecules
                 and  a weight arising from 
                     thermodynamic stability. 
                     Also, roughly included is   
                     the effect of the structure
                      of the interface 
                     which is  
                     anisotropic and it is known 
            to be relevant factor\cite{rozmanov2012anisotropy,espinosa2016time}.
            In this work, we address  all these
            three factors. We compute the 
            diffusion and the thermodynamic stability from bulk
             simulations of ice and water, and we investigate 
             the surface anisotropy by comparing the kinetics
              of   three different facets for the ice-water
               interface. Moreover, 
            we investigate the effect of pressure which
             is known to play an important role in ice nucleation\cite{espinosa2016interfacial, bianco2021anomalous}.
            Hence, we 
               reach for the first time ab initio accuracy
               in the description of the kinetics of the
                ice-water interface including multiple factors.\\
                
               This article is organized as follows. First,
               we present the methods, including simulation details.
                Then, we show the results section which includes three subsections. The first one covers ice growth and melting rate
                 from molecular simulations
                 and the effect of the
                 structure of the interface.
                 Then, 
                 the comparison with   experiments
                   and the Wilson-Frenkel  framework follows.
                    Finally we show the
                     effect of pressure in both ice growth
                      and melting. At the end, the
                      conclusions are presented.

\section{Methods \label{sec:methods}}

\begin{figure}[h!]
\centering
\includegraphics[width=2.8in]{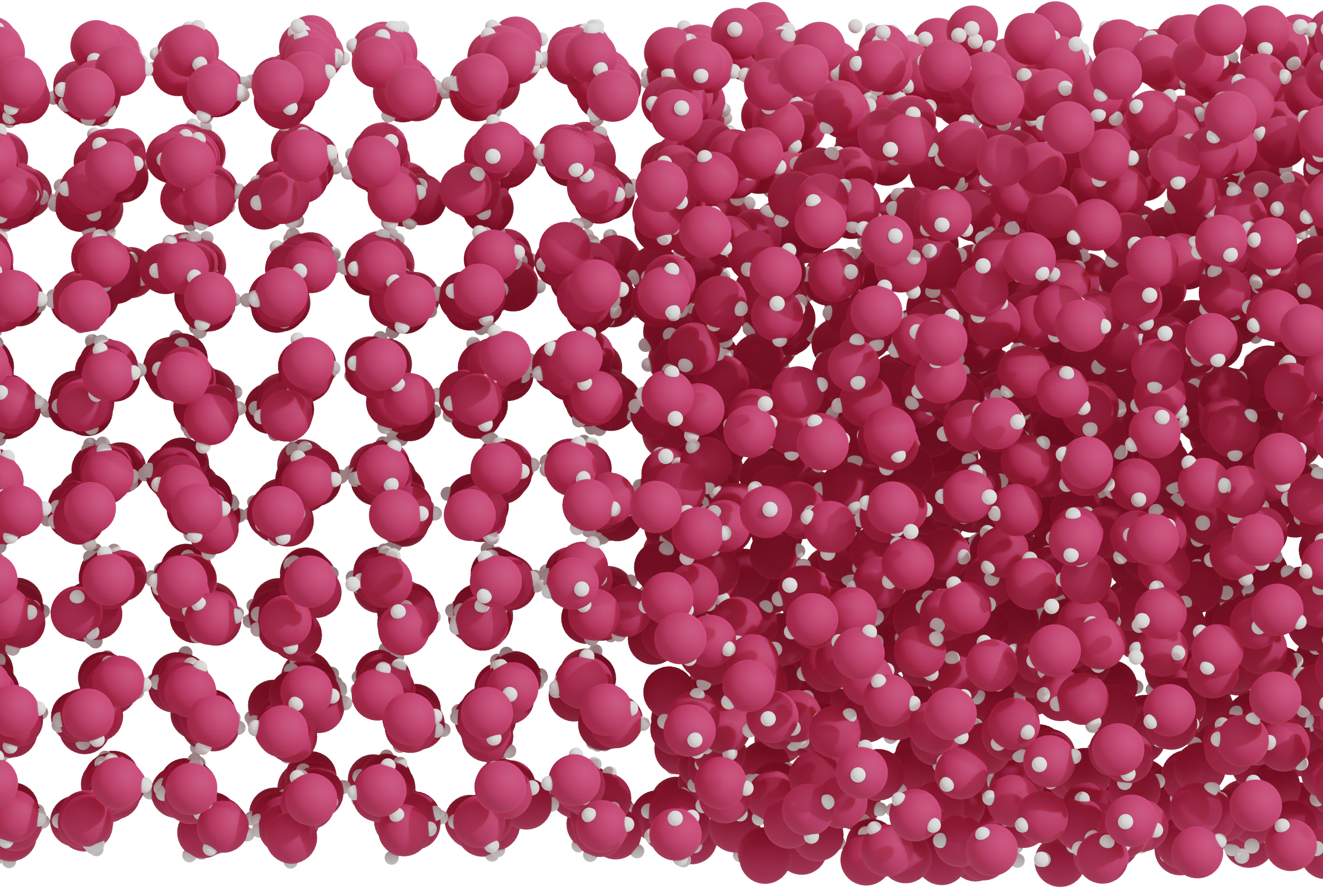} 
\caption{\label{fig:setup} Ice-water interface through the
 basal plane. }
\end{figure}

\begin{figure*}[]
\centering
\includegraphics[width=1.5in]{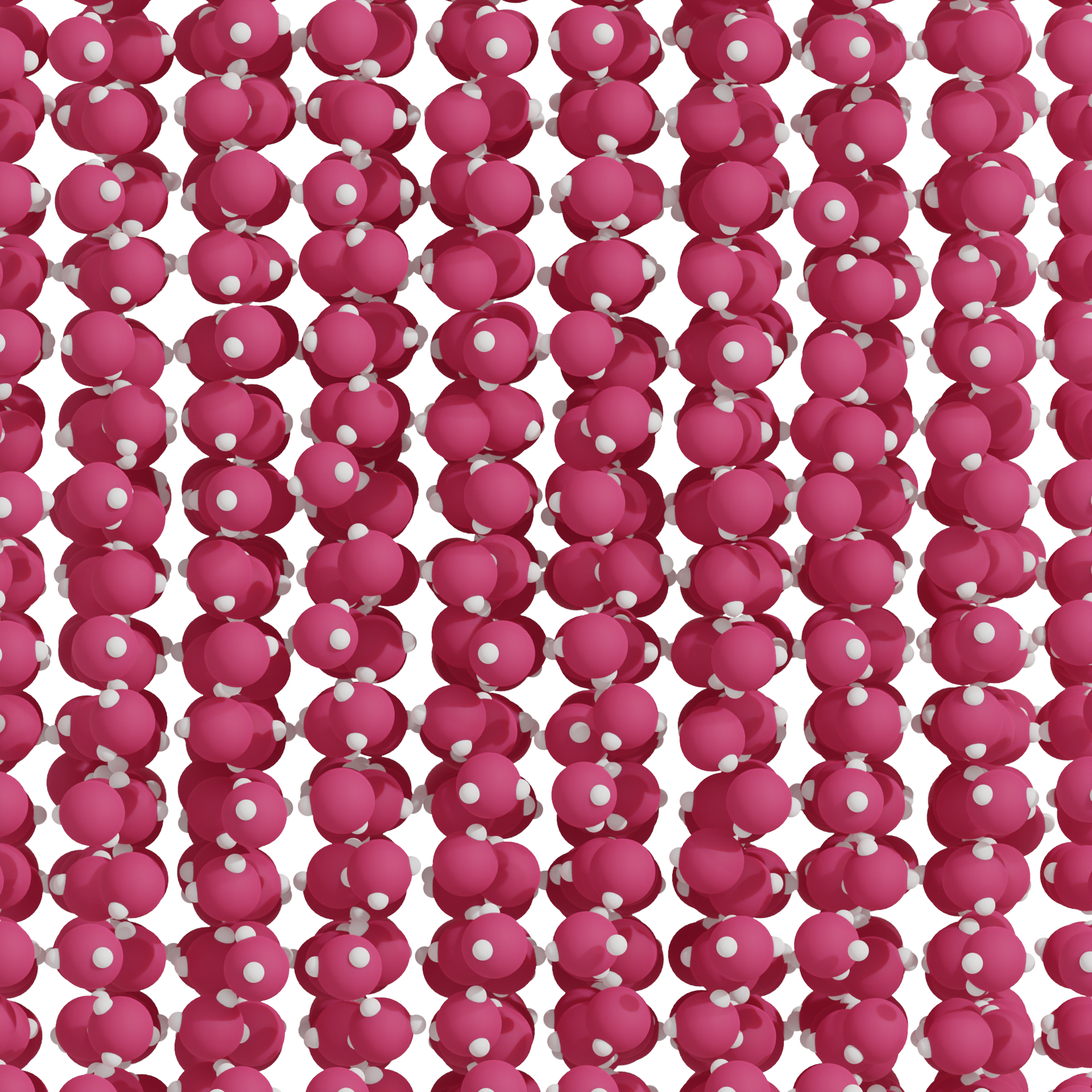} a)
\includegraphics[width=1.5in]{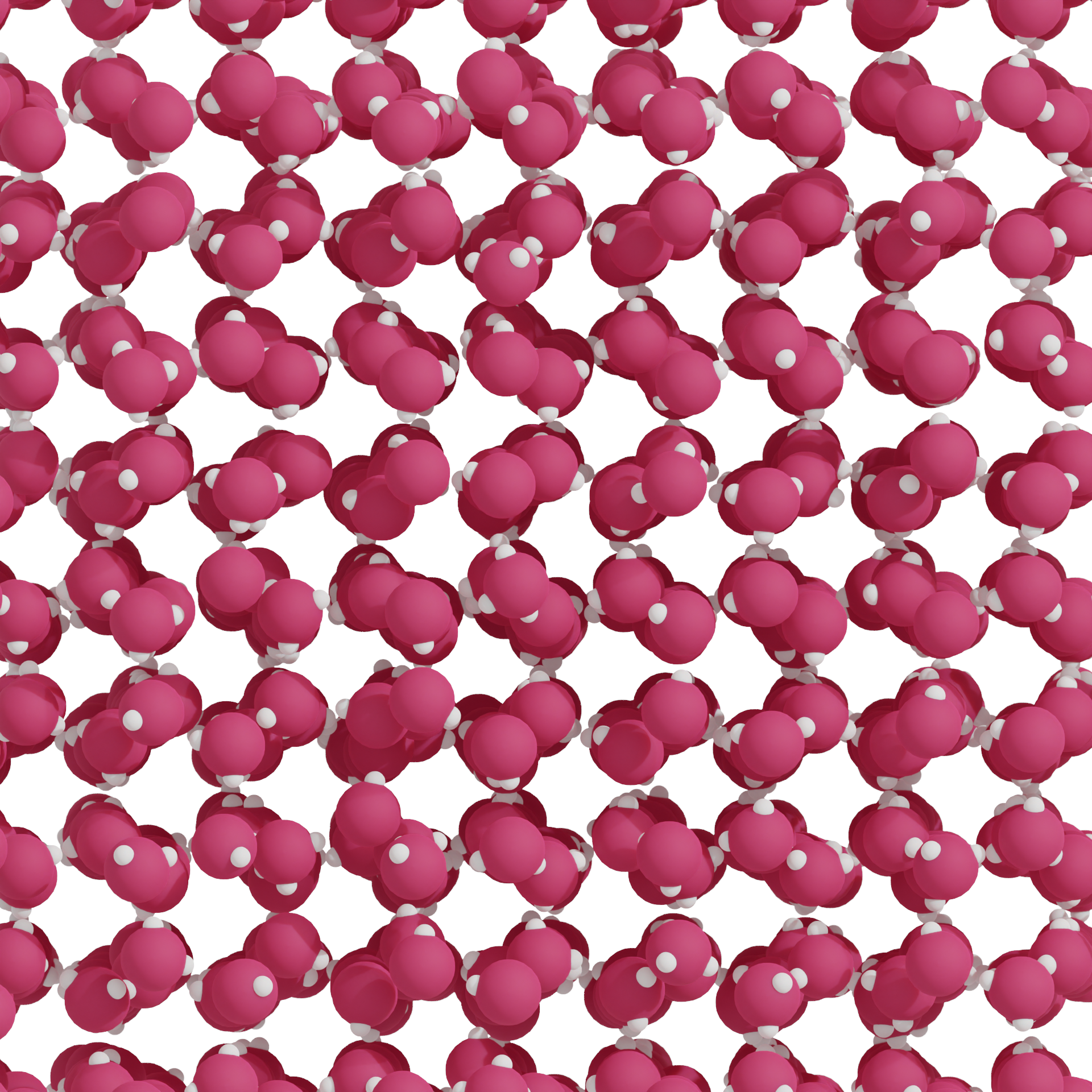} b)
\includegraphics[width=1.5in]{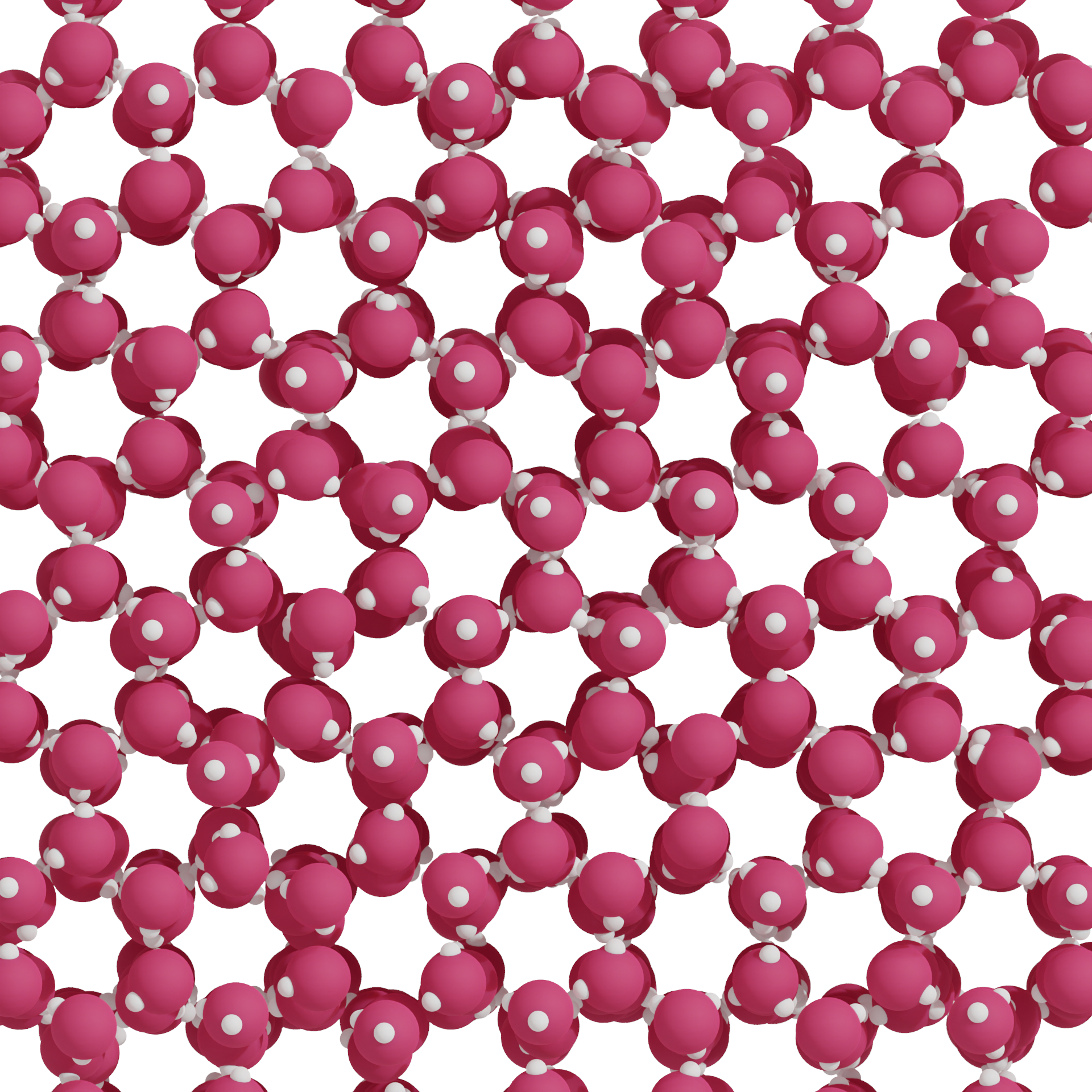} c)
\caption{\label{fig:setup2} From left to right: the  
 structure of the primary prismatic, secondary prismatic,
  and basal facets.}
\end{figure*}

 In this work, we  adopt the water model proposed in Morawietz et al.\cite{morawietz2016van} which is based
  on the Behler-Parrinello approach\cite{behler2007generalized}. This is, we describe water interactions 
 with a  
 neural-network potential  (NNP). In Ref. \cite{morawietz2016van},
  the NNP was    
 trained on  
 projector-augmented-wave (PAW) 
 calculations  using the 
  Revised Perdew–Burke–Ernzerhof (RPBE) \cite{hammer1999improved} functional including
  van der Waals interactions with the D3 method \cite{grimme2010consistent}.
  We use 
the open-source LAMMPS package \cite{plimpton1995fast} 
to run the  molecular dynamics
  simulations. In particular, we employ the
  n2p2 extension that allows for the usage of
   NNP's\cite{singraber2019parallel}.
  All simulations are performed with a timestep of 0.5 fs. We employ the
   Nose-Hoover \cite{nose1984unified} thermostat and barostat with relaxation times  of
   0.05 ps and 0.5 ps respectively. 
 The setup to study the kinetics of the
  ice-water interface consists of a slab of ice in contact with a slab  of water
 as can be seen in Fig. \ref{fig:setup}. Thus,
  we use direct coexistence simulations\cite{garcia2006melting}.
  Ice is prepared with the GenIce package\cite{matsumoto2018genice}. The
    systems have about 3500 molecules in total. The 
  interface is always parallel to one
   of the Cartesian planes exhibiting the
   basal, the primary or the secondary prismatic
    facet which are shown 
     in Fig. \ref{fig:setup2}. \\
 
\begin{figure}[h!]
\centering
\includegraphics[width=3.1in]{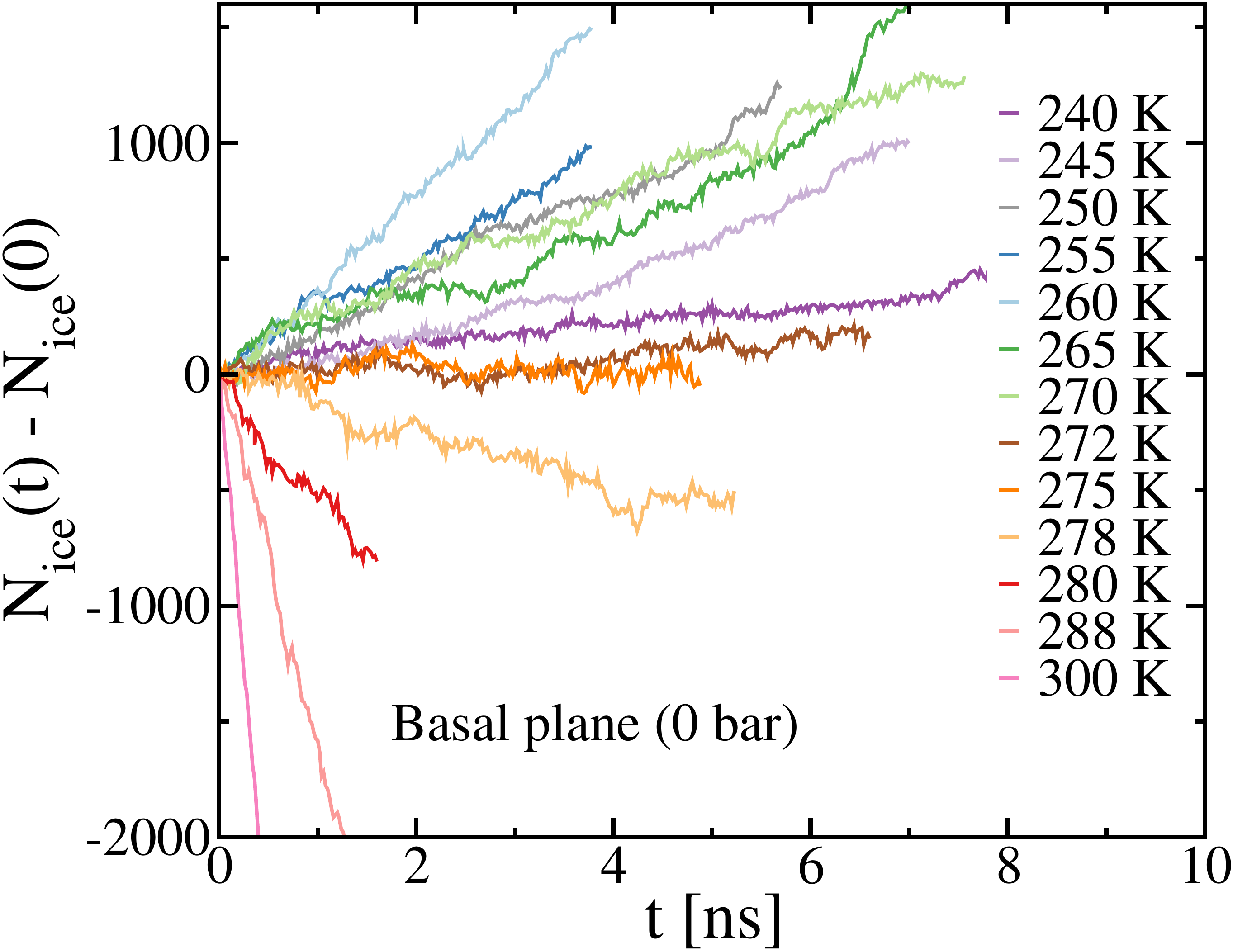} 
\caption{\label{fig:basal} Change in number of ice molecules in the 
 ice block over time.  The ice plane is basal and the pressure is 0 bar. Following the legend, 
 color  in the
  line of the set indicates the temperature $T$. 
  The melting
   temperature $T_{m}$ (274 K in this case) is found from the average in $T$ 
   between two
    sets having
    different sign in the slope of 
    $N_{\rm ice}(t) - N_{\rm ice}(0) $ that are consecutive in $T$.  
  }
\end{figure}

This system is simulated in the anisotropic isobaric-isothermal ensemble (NpT) at different
 temperatures from 240 K to 300 K along the -1000, 0, and 2000 bar  
  isobars. Even though ice melting or growth involves latent heat, one can apply thermostats as shown
   in Ref.\cite{montero2019ice}. Periodic
   boundary conditions are used so that two interfaces are exposed.
 During the simulation,  ice grows, melts, 
 or remains the same size depending on temperature $T$
  and pressure $p$. 
   We can
   observe this by computing the change in the number of ice
    molecules $N_{\rm ice}(t) -  N_{\rm ice}(0)$ in the system as a function of time
    as shown in Fig. \ref{fig:basal}.
    In order to obtain $N_{\rm ice}(t)$
    we use the order parameter
     proposed by Lechner and Dellago\cite{lechner2008accurate} 
     to identify ice-like  and  
     water-like molecules. 
 We compute $N_{\rm ice}(t)$ before the two interfaces are
      too close since they may interact from a certain distance
      due to interface 
  fluctuations\cite{moritz2020weak}. 
    In fact, the slope of $N_{\rm ice}(t)$ 
    is used to 
    obtain the speed of the ice-water interface $u$
    defined as
    
\begin{equation}
u = \frac{1}{2}\frac{r(t)-r(0)}{ t} .   
\label{eq:uu}
\end{equation}

Here, $r(t)$ is the thickness of the slab  
at time $t$ given as 

\begin{equation}
    r (t)= \frac{N_{\rm ice}(t)}{\rho_{\rm ice}A},
\end{equation}   

\noindent where  $\rho_{ice}$  is the average ice density and $A$ is the average interfacial area which are constants.  The division by 
 2 in Eq. \ref{eq:uu} is necessary because of the periodic boundary conditions and so the ice slab grows/shrinks on both of its sides. Finally, one finds the linear relation
   between $N_{\rm ice}$ and $t$, 

\begin{equation}
     N_{\rm ice}(t) -   N_{\rm ice}(0) = 2A\rho_{\rm ice}ut ,
\end{equation}
 in agreement with the results shown in Fig. \ref{fig:basal}. 
In order to estimate the uncertainty in  $u$, we compute by
  block averaging the error of the slope of $N_{ice}(t)$ and neglect the errors of $A$ and $\rho_{ice}$. 
 \\

We also conduct simulations of both bulk ice
 and bulk water. We use NpT parallel tempering at each isobar 
 for 1 ns to
  ensure the equilibration of 
  water\cite{okabe2001replica,mori2010generalized} before running
  the standard NpT simulation for 4, 8, and 16 ns for the 2000,
   0, and -1000 bar isobar, respectively. This 
  allows us to estimate
  the diffusion coefficient $D$ and the difference
   in chemical potential $\Delta \mu $ 
   between the phases
    at a given thermodynamic state.  We compute the mean square 
     displacement as a function of 
     $t$ in bulk water and then
     we find $D$ from the relation
     
     \begin{equation}
        \langle | \vec{r}(t) - \vec{r}(0)  |^2   \rangle = 6Dt.
     \end{equation}
     
 It is well known that $D$ suffers
  from finite size effects\cite{montero2018viscosity,morawietz2016van}.
  For our system size, the expected error of $D$ is less than $5\%$.
    $\Delta \mu $ 
  is obtained
 by thermodynamic integration from coexistence along the
  isobar of interest for both ice and 
 water\cite{vega2008determination}, 

 \begin{equation}
        \bigg |   \frac{\Delta \mu}{k_{B}T}  \bigg | =   \bigg |  \int_{T_{m}}^{T} \frac{1}{k_{B}T^{'2}}\left( \frac{H_{ice} }{N_{ice}}  - \frac{H_{w}}{N_{w}}\right)dT^{'}  \bigg | .
       \end{equation}
       
 The melting temperature $T_{m}$ is obtained
  from the direct coexistence simulations
   of ice and water\cite{garcia2006melting} where
    we also measure $N_{ice}(t)$. 

\section{Results \label{sec:results}}

  \subsection{Ice growth rate from molecular  simulations}

\begin{figure}[h!]
\centering
\includegraphics[width=3.4in]{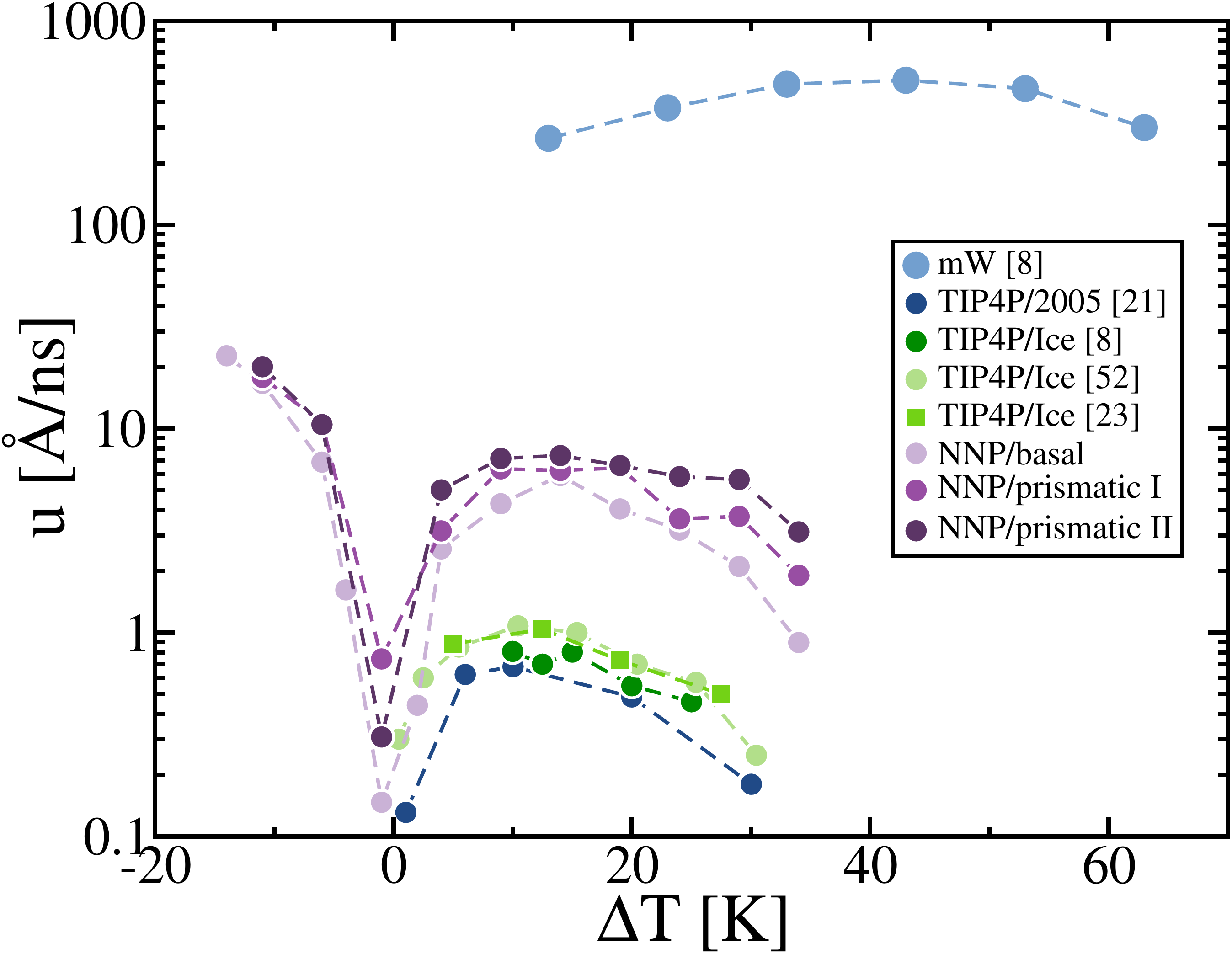} 
\caption{\label{fig:icegrowthsims} The speed of the
 ice-water interface $u$ is shown 
for different 
water models as a function of  $\Delta T = T_m - T$. 
 Data for mW (light blue circles) are  taken from Ref. \cite{espinosa2016time} and 
  for TIP4P/2005 (dark blue circles) 
  are from Ref. \cite{rozmanov2011temperature}.
  Two sets for thermostated simulations of
   TIP4P/Ice are included 
   shown as  light green\cite{weiss2011kinetic} and dark green circles\cite{espinosa2016time}.
    In green squares, we show data from
     Ref. \cite{montero2019ice} for TIP4P/Ice
     in the NVE ensemble, i.e. without 
      thermostat. The results from the NNP cover also the overheating
       regime, i.e. $\Delta T < $ 0. 
       These are shown as purple circles,
       light for basal, middle for prismatic I,
        and dark for prismatic II. The dashed lines are guides for
         the eyes. 
  }
\end{figure}

Here, we compare our results for 
     the ice growth rate for the NNP with
     those of empirical force fields
      including   TIP4P/2005\cite{abascal2005general}
      from Rozmanov and Kusalik\cite{rozmanov2011temperature},   TIP4P/Ice\cite{abascal2005potential} from Refs. \cite{espinosa2016time,weiss2011kinetic,montero2019ice}, and  mW\cite{molinero2009water} from Espinosa et al. \cite{espinosa2016time}. 
      In  Fig. \ref{fig:icegrowthsims}, the speed of the ice-water interface
       $u$
       is shown as a function of $\Delta T = T_{m} - T$.    First, we focus our analysis to the supercooled
   regime, i.e. $\Delta T > 0$.
       For the NNP, 
        we present the results for the basal and primary and secondary prismatic facets.
      As can be seen in Fig. \ref{fig:icegrowthsims},
       the qualitative behavior is the same 
       for all models with an initial increase of the
        growth rate followed by a
         maximum and then a decrease.
       However,
       the ice growth rate from the 
       NNP is almost one order of magnitude
        larger than 
         TIP4P/Ice and TIP4P/2005 
         although almost two 
          orders of magnitude
          smaller than  mW. 
          The mW is a monoatomic water
           model\cite{molinero2009water}, thus, it lacks the 
            rotational and vibrational motion
            while the TIP4P/Ice\cite{abascal2005potential} and TIP4P/2005\cite{abascal2005general}
        are 4-site rigid models so that they
        can have rotational but no vibrational
        motion. 
         The NNP\cite{morawietz2016van}
         has no such constraints and
          atoms can, in principle, move
           freely. 
         Thus, this model allows molecules to
          display any kind of motion. 
          Further work is needed to isolate
           the effect that each degree of 
           freedom in the water molecule
           has over the ice growth rate. 
             In any case, even though the ice-water
              interface of NNP
              is faster than that of 
               TIP4P/Ice and TIP4P/2005,
                the maximum in the ice growth
                 rate is located 
                  at the same supercooling 
                within the uncertainty. 
                This is, $\Delta T = $ 14 $\pm$ 1 K. 
                In the case of the NNP, the maximum ice growth rate 
                  $u_{max}$ is 6.5 $\pm$ 1 $\text{\AA}$/ns.\\

     In order to assess the effect that the exposed facet 
   may have
  on the kinetics of the ice-water interface,
  we investigate  the basal, 
  the primary prismatic,
   and the secondary prismatic facets including
    also  the
    overheating regime, i.e. we discuss the ice melting rate. 
    In contrast to ice growth,
the kinetics of melting seems to be monotonic with $\Delta T$
 for any facet.  
   Furthermore, we observe that both prismatic
     facets present similar kinetics (although the secondary prismatic
      face is faster
     for large supercoolings), 
     which is 60$\%$ faster than that of the basal plane. 
         This is in agreement
          with what was observed for  TIP4P/Ice\cite{weng2022investigation,espinosa2016time,espinosa2016ice} and TIP4P/2005\cite{rozmanov2012anisotropy}.

 \subsection{Comparison with Wilson-Frenkel theory and  experiments}
 
\begin{figure}[h!]
\centering
\includegraphics[width=3.4in]{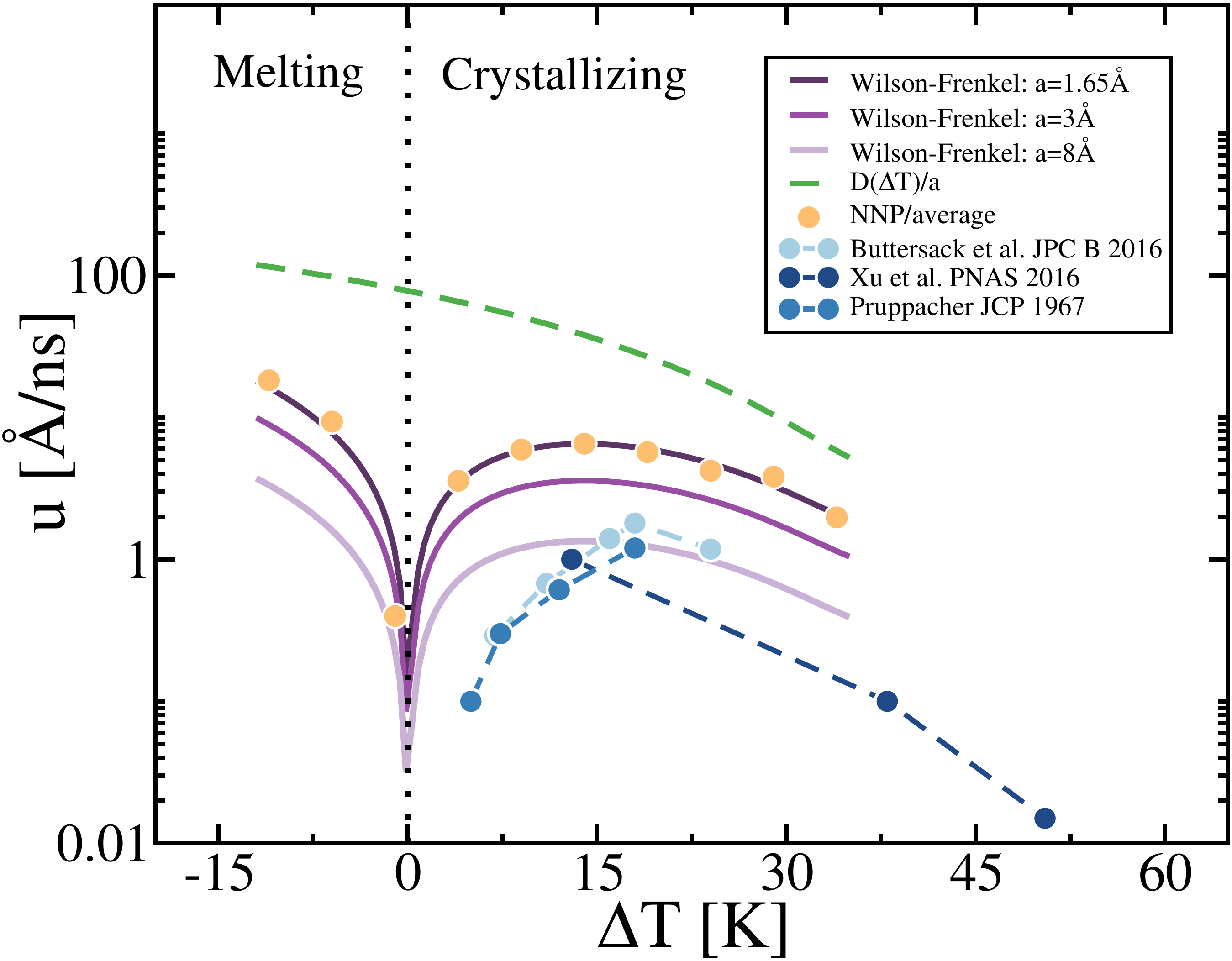} 
\caption{\label{fig:icegrowthexp} The ice growth rate from the NNP averaged over basal and 
prismatic I and prismatic II facets (orange circles) compared 
 to experiments (dark to light blue circles connected by dashed
  lines) \cite{xu2016growth,buttersack2016critical,pruppacher1967interpretation}.  Also shown
  are Wilson-Frenkel curves for different
   values of the characteristic length $a$ (dark to light purple lines),
    and the case of diffusion limited 
    growth/melting for $a= $ 1.65  $\text{\AA}$ (green line). }
\end{figure}

 The Wilson-Frenkel relation\cite{wilson1900xx,frenkel1932note} provides an expression for crystal growth 
 that takes into account
    both the thermodynamics and the kinetics
     of the system. It is obtained
      from the attachment and detachment rate
       of molecules in the interface\cite{tepper2001molecular}. 
     The general expression involves several parameters
      but it can be approximated by 
       the following relation,
 
 \begin{equation}
    u(T) = \frac{D(T)}{a}\left[ 1 - \exp \left(-\frac{|\Delta \mu (T)|}{k_{\rm B}T}\right)\right],
    \label{eq:WF}
\end{equation}

 \noindent where $a$ is a characteristic length\cite{montero2019ice,garcia2006melting,xu2016growth} which is
   often set equal to a molecular
   diameter \cite{montero2019ice,garcia2006melting} ($\sim $3 $\text{\AA}$ in
    the case of water).
   However, in Ref. \cite{xu2016growth}, $a$ was estimated to be
    8  $\text{\AA}$ 
    from Eq. \ref{eq:WF} providing experimental values to $u$, $D$,
     and the thermodynamic driving force.
     Eq. \ref{eq:WF} states that $u$ is proportional to 
  $D$ rescaled by $a$ and weighted by the thermodynamic 
  stability $\Delta \mu$.
In Fig. \ref{fig:icegrowthexp}, we show
  $u(T)$ as obtained from Eq. (\ref{eq:WF}).
  For $D$ and $\Delta \mu$ we use 
   values 
  from equilibrium simulations of
   bulk ice and bulk water at 0 bar. We find that the
   best fit to our NNP simulations  occurs under 
 $a = $ 1.65  $\text{\AA}$,
 whereas with $a = $ 8  $\text{\AA}$
  we are closer to experimental results. 
    In Fig. \ref{fig:icegrowthexp},
    we  also include, $u$ from the NNP averaged over the three facets
     as well as experimental results
       for the ice growth rate. 
       The laser-pulsed nanofilms provided
       by Xu et al.\cite{xu2016growth} allowed
       for measurements at very deep supercoolings
       with high control of temperature 
       proving experimentally the decay in the 
       interface growth rate with
        increasing the supercooling. For lower
         supercoolings, however, experiments
           show an increase\cite{buttersack2016critical,pruppacher1967interpretation} so that
          a maximum is also expected around $\Delta T =$ 18 K 
          in reasonable agreement with the simulations
           and the Wilson-Frenkel theory.
           Some factors that may explain 
           discrepancies between experiments
           and our simulations are the role
            of impurities (they slow down
    the kinetics)\cite{carignano2006molecular,bauerecker2008monitoring,wahl2020ice,lin2020interaction}, ambiguous determination of temperature during a phenomenon
             involving heat exchange\cite{meng2020dynamic},
              and the fact that, in experiments,
              ice can grow in dendritic form or in
               layers under certain
                circumstances
                which may produce growth rates
                differing in about two orders
                 of magnitude for the same
                  experiment\cite{kapembwa2014heat}. Also within  Fig. \ref{fig:icegrowthexp}
                can be seen that 
               even
    though the Wilson-Frenkel is a model for  crystal growth, it describes surprisingly 
    well the kinetics of melting in the moderate overheating regime showing a monotonic increase with
     overheating. Therefore, the diffusion coefficient
      $D$ and the thermodynamic stability represented
       as $\Delta \mu$ play a crucial role.
       In order to separate these effects, we compute
       the diffusion limited kinetics, i.e. we compute
        $D(\Delta T)/a$. As can be seen,
        the thermodynamic stability dominates
        for a small degree of metastability,
         whereas for large supercooling
          diffusion almost enterily dominates.\\

\subsection{The role of pressure in ice growth and melting}
 
\begin{figure}[h!]
\centering
\includegraphics[width=3.1in]{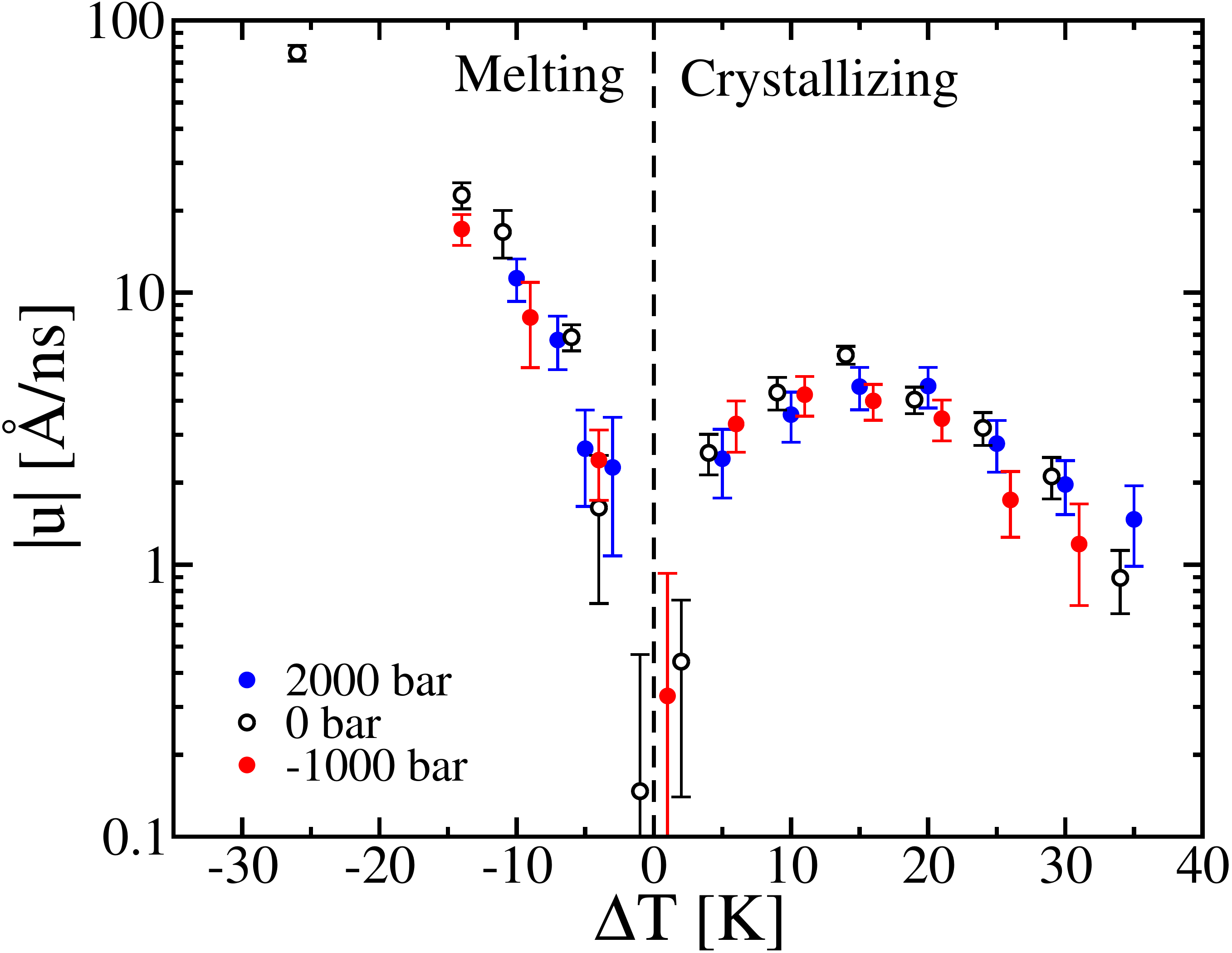} \caption{\label{fig:pressure} Ice growth/melting rate as a function of $\Delta T = T_{m} -T$ along different isobars. The exposed plane is the basal. }
\end{figure}

\begin{figure}[h!]
\centering
\includegraphics[width=3.1in]{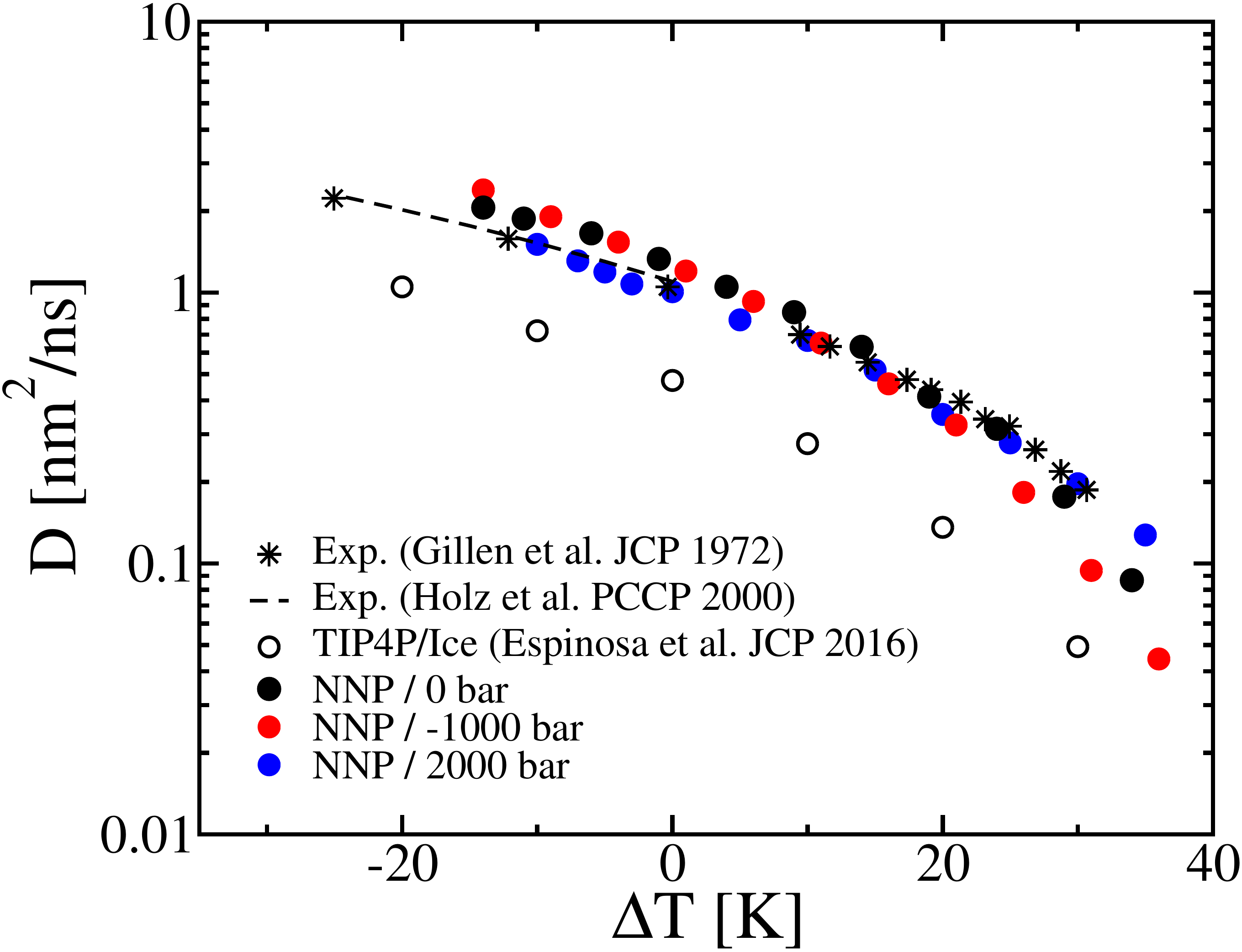} a) 
\includegraphics[width=3.1in]{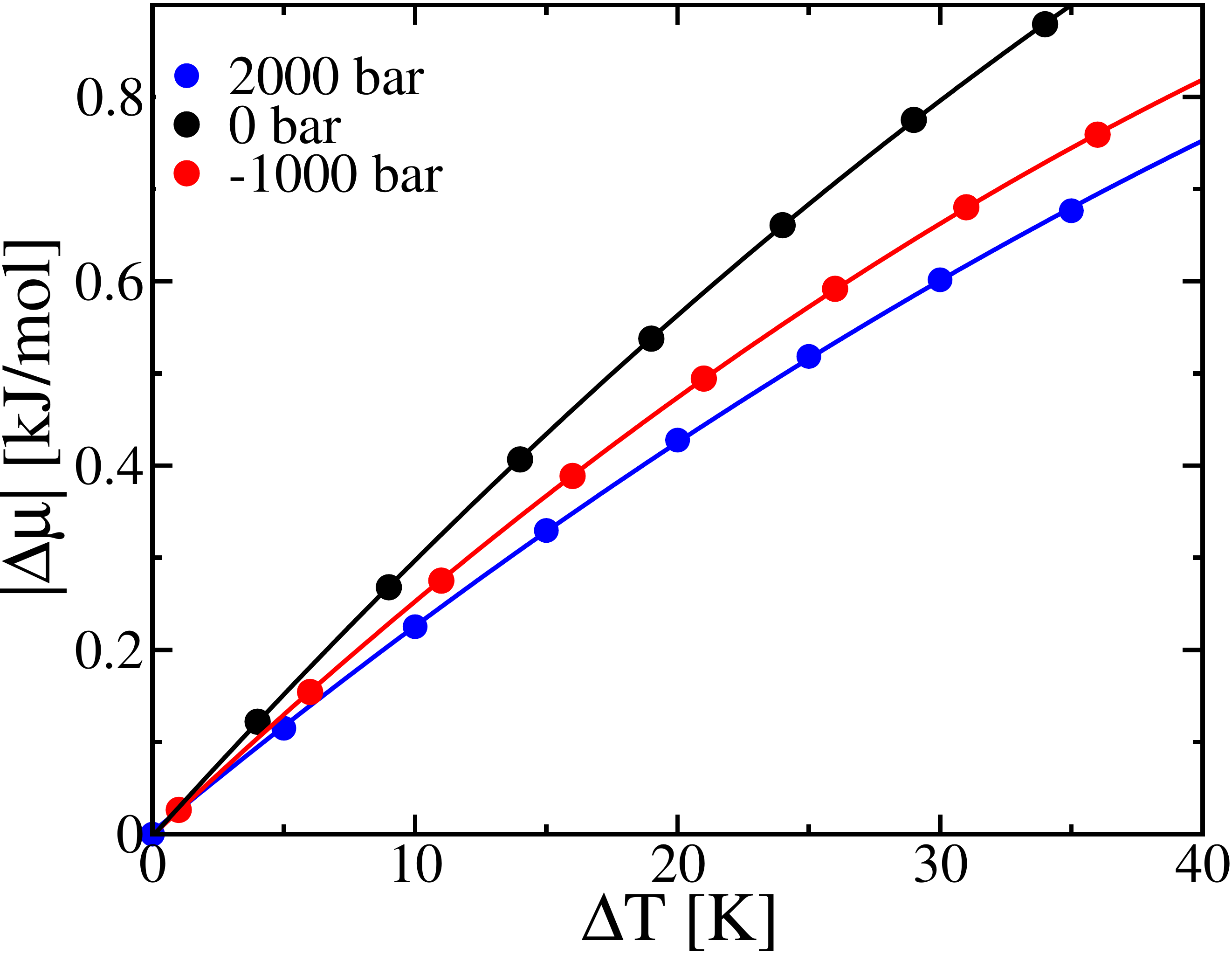} b) 
\caption{\label{fig:pressure2} 
 a) Diffusion
  coefficient against $\Delta T = T_{m} - T$ for three 
   different isobars (0 bar in black, -1000 bar in blue,
    and 2000 bar in red). 
  Experimental results are also included  \cite{holz2000temperature,gillen1972self}. 
  b) Difference in chemical potential $\Delta \mu$ against $\Delta T$
   along the same isobars as in panel b).}
\end{figure}

 In order to assess the effect that pressure may have in the growth rate of ice, we study three isobars. Apart from the 0 bar
  isobar, we include here 
  one isobar at negative pressure (-1000 bar) so that the system is
  stretched and one at 
   high pressure (2000 bar) where the system is
    compressed. We obtain
   $T_{m}^{(-1000)} = $ 276 K, 
   $T_{m}^{(0)} = $ 274 K,
   and $T_{m}^{(2000)} = $ 265 K. As can be seen
   in Fig. \ref{fig:pressure},
    when the growth rate is presented against $\Delta T$ 
     the effect of pressure is very small both
      in the melting and crystallization regimes.
      Only at high supercoolings a small deviation
       seems to emerge.
      According to the 
    Wilson-Frenkel model, $D$ and $\Delta \mu$
     are the main factors determining $u$.
   As can be seen in Fig. \ref{fig:pressure2} a) 
  for the diffusion coefficient $D$  and in 
  b) for $\Delta \mu$ both as a function
   of supercooling, the pressure dependence is very
    weak from small to moderate supercoolings. 
    For higher supercoolings, a small effect
     of pressure 
    in $D$ and $\Delta \mu$ is seen. Let's compare
      the negative and high pressure isobars
       with the standard pressure one. 
       On the one hand, 
     $D$ is larer at 2000 bar in this regime
      but also $\Delta \mu$ is smaller. Thus,
       the faster kinetics is hindered by a smaller
       driving force. On the other hand, 
     both  $D$ and $\Delta \mu$ are slightly smaller 
     at -1000 when compared to the 0 bar isobar.
    Looking back to Fig. \ref{fig:pressure},
     the -1000 bar isobar seems to have slightly
      slower kinetics at high supercooling
       in agreement with the slightly lower 
    $D$ and $\Delta \mu$ albeit still
     within the error bars of the other
      isobars. Therefore, $u$ as a function of $\Delta T$
       is roughly independent on the isobar. This 
       resembles to something recently observed in
        ice nucleation with the TIP4P/Ice model\cite{de2022minimum}.
     There, it was shown that the critical nucleus size, the 
      interfacial free energy, the free energy barrier, and 
       the nucleation rate  are  roughly independent of the
         isobar as long as pressure is between -2600 bar
          and 500 bar. Hence, it seems quite general, in
           both ice nucleation and growth, 
           that supercooling is a way to roughly map different isobars
     of a certain variable to the same curve. 
     In principle, this should be the case
       as long
      as the melting temperature does not change significantly with 
      pressure.\\

 Finally, in  Fig. \ref{fig:pressure2} a), 
       we can see that $D$ as obtained from the
        NNP has a better agreement with experiments\cite{gillen1972self,holz2000temperature}
         than  TIP4P/Ice. This is in contrast
          with the result of the ice growth
          rate where the agreement with 
           experiments was better
           for  TIP4P/Ice.
         Furthermore,   by considering the diffusion coefficient as a function
  of temperature instead of supercooling, one could see that,
   within the studied range of temperatures and pressures, 
   larger pressure corresponds to larger $D$. 
  The existence of this anomalous diffusion in water along isotherms,
   within a certain pressure regime, is well known from empirical force fields\cite{montero2018viscosity} and experiments\cite{prielmeier1988pressure}. Here, we show that it also
    emerges from  an ab initio machine learning force field.\\

\section{Conclusions}

 In this work, we have investigated the kinetics of the ice-water interface by means
    of an ab initio machine learning approach.
    In particular, we have used the neural-network
     potential proposed in Ref.\cite{morawietz2016van} 
     to reach ab initio accuracy
      for a system of about 3500 molecules simulated
       for long times (more than 500 ns in total). 
  Comparison of our results for the ice growth rate with
   previous simulations and experiments shows reasonably good 
    agreement.
    The maximum ice growth rate, 6.5 $\text{\AA}$/ns, is achieved at 14 K below
     the melting temperature.
     The dependence of the growth rate can be explained in terms of
     the Wilson-Frenkel theory, according to which
      the kinetics 
      depends mainly on the mobility of molecules and
       the relative thermodynamic stability of the phases.
 %      Near the melting point, the thermodynamic stability plays
 %       an important role,  whereas far from the melting,  
  %      diffusion is almost completely dominant. 
       The structure of the surface also plays an important
        role, as we have quantified by studying different
         surfaces, i.e. the basal and the primary and
          secondary prismatic facets. The prismatic
           facets grow at a similar rate and about 60$\%$ faster than
            the basal facet. 
         In contrast to  ice growth, 
         the kinetics of melting is found to be
     monotonic
      with the degree of  overheating. This is because both diffusion and driving force do not compete with each other but 
       work in the same direction in this case. We studied 
      the kinetics of the interface along three isobars, 
      at -1000, 0, and  2000 bar, for which we obtained the melting points
        276, 274, and 265 K respectively. However, pressure is found
       to be barely relevant when the kinetics is studied
       against supercooling or overheating.
        %Nevertheless, we suggest that this is
      % due to the combined action of mobility,
       %  driving force, and surface structure.
         %The effect
         % of the surface at different pressure is inferred
         %  from the interfacial free energy which is
         %   approximated from Turbull's heuristic relation.
         We observe that
             the anomalous diffusion of water is also captured by the neural-network potential, i.e., diffusion increases upon
        increasing the pressure. 
      Therefore, a complete description of the kinetics of the ice-water interface
       including the effect of several factors and under different isobaric conditions is provided
        from an ab initio machine learning approach to molecular simulations.

\section{Acknowledgments}

The authors acknowledge the support from the  SFBTACO (project nr. F81-N) funded by FWF as well as 
 the computer resources and technical assistance provided by the Vienna Scientific Cluster (VSC).

\section{Author declarations}

\subsection{Conflict of Interest}

The authors have no conflicts to disclose. 

\subsection{Data availability}

The data that support the findings of this study are available
from the corresponding author upon reasonable request.

\bibliographystyle{ieeetr}
\bibliography{newbib}

\end{document}